\documentclass[aps,prb,twocolumn]{revtex4-2}
\usepackage{color}
\usepackage{graphicx}
\usepackage{cancel}
\usepackage{array}
\usepackage{amsfonts}
\usepackage{amssymb}
\usepackage{amsthm}
\usepackage{amsmath}
\usepackage{appendix}
\usepackage{url}
\usepackage{hyperref}
\usepackage{float}
\usepackage{amsmath}
\usepackage{multirow}  % multirow tabular
\usepackage{dcolumn}   % Align table columns on decimal point
\usepackage{bm}        % bold math
\usepackage{enumerate}
\usepackage{mathtools}
\usepackage{tabularray}

\usepackage[dvipsnames]{xcolor}
\hypersetup{
	colorlinks,
	linkcolor={green!50!black},
	urlcolor={blue!80!black},
	citecolor   = {red!50!black} %Colour of citations
}

\def  \bsig    {\mbox{\boldmath$\sigma$}}
\def  \bD    {\mbox{\boldmath$\Delta$}}

\def  \btau    {\mbox{\boldmath$\tau$}}

\begin{document}

\title{Longitudinal magnetoresistance in graphene with random Rashba spin-orbit interaction}

\author{S. Kudła$^{1}$, 
%V. I. Ivanov$^{2}$, 
V. K. Dugaev$^{1}$,
%E. Ya. Sherman$^{2}$,
J. Barna\'s$^{2,3}$, A. Dyrda\l$^{2}$}

\email[e-mail: ]{adyrdal@amu.edu.pl}
\affiliation{$^{1}$Department of Physics and Medical Engineering, Rzesz\'ow University of Technology,
35-959 Rzesz\'ow, Poland\\
%$^{2}$Department of Physical Chemistry, Universidad del Pa\'is Vasco UPV/EHU, 48080 Bilbao, Spain
$^{2}$Faculty of Physics, ISQI, Adam Mickiewicz University in Pozna\'n, ul. Uniwersytetu Pozna\'nskiego 2, 61-614 Pozna\'n, Poland\\
$^{3}$Institute of Molecular Physics, Polish Academy of Sciences, ul. Smoluchowskiego 17, 60-179 Pozna\'n, Poland\\
}

%\ESc{This font is for suggestions from E.S.}

\date{\today}

\begin{abstract} 
We consider longitudinal electronic transport in a graphene monolayer, with an external in-plane magnetic field parallel to the electric field, and in the presence of extrinsic spatially fluctuating spin-orbit Rashba interaction. Our main interest is focused on the longitudinal magnetoresistance. We show, that for the spin-orbit correlation length $\xi$ between $10^{-6}$ and $10^{-5}$ cm, scattering processes on the Rashba spin-orbit fluctuations lead to a negative magnetoresistance. For smaller correlation lengths,  $\xi <10^{-6}$ cm, we find a small and positive magnetoresistance. The description is based on an effective model of graphene Hamiltonian, that is valid for low energy states around the Dirac points. In turn, to derive the electron scattering rate and the longitudinal conductivity and magnetoresistivity, we employ the linear response theory and Green's function formalism. 
\end{abstract}

%\pacs{}

\maketitle

\section{Introduction}

Graphene-based van-der-Waals heterostructures seem to be an excellent platform for future two-dimensional (2D) electronics and spintronics~\cite{Fabian_rev2014,Valenzuela2022,Kurebayashi2022,Sierra2021,Inoue_book}. With its low-energy linear dispersion as well as valley and pseudospin degrees of freedom, graphene ensures better functionality of new generation electronic and logic devices~\cite{CastroNeto_RevMOdPhys2009,falkovsky07,Roche_book,Katsnelson_book}. Moreover, the magnetic and spin-orbital proximity effects in graphene deposited on other van-der-Waals  crystals  make it useful for  spin-to-charge or charge-to-spin  conversion phenomena,  and thus also for generation of spin-orbital torques~\cite{Dyrdal2015,milletari16,wang16,Dyrdal_2017,offidani18prb,offidani18prl,Garcia_2019,Benitez_2020,Roche_PRL2024,perkins24,Wojciechowska_2024,Fert_RevModPhys2024}.

When considering applications, full understanding and control of transport phenomena, not only in pristine graphene, but also in disordered van der Waals structures,  are highly desired.  The effects induced by various kinds of disorder in electronic transport properties of graphene and graphene-based hybrid structures are intriguing and widely discussed in the relevant literature. As a result of single 2D crystal preparation (e.g., on the stage of graphene exfoliation or various growth methods) and fabrication of van-der-Waals heterostructures and devices (transfer to other substrate or twist with respect to other layers in heterostructure, contacts deposition, etc.), one can expect long- or short-range imperfections, like ripples, impurities, vacancies, or strain in the system~\cite{Inoue_book,Roche_book,arndt09,Cresti2008,offidani18prb,ferrell16,elias17,alekseev13,uppstu12,rappoport09}. In addition, the co-existence of various kinds of disorder in systems with linear low-energy spectrum  and coupling between spin and pseudo-spin degrees of freedom leads to  various -- sometimes unexpected -- contributions to the electronic and spin relaxations, and thus to   conductivity~\cite{Hwang2009,Han2012,Zhang_2012,fabian16,jabakhanji14,narozhny15,soriano15,chiappini16,cummings17,garcia17,Ringer2018,cummings18}. 
For instance, quantum corrections to the electrical conductivity are examples of such effects~\cite{mccann06,jobst12,chiappini16,sousa22}. 

It is well-known that in 2D electron gas without spin-orbit coupling (SOC), one can observe  the weak localisation (WL) phenomenon as a consequence of constructive phase interference in electron self-crossing trajectories. However, in the presence of Rashba spin-orbit coupling, the weak anti-localization (WAL) appears as a result of destructive interference of electron trajectories due to the spin dynamics induced by SOC. In turn, in case of pristine graphene (no Rashba SOC), the presence of an additional degree of freedom, i.e. pseudospin, leads to  WAL, whereas when Rashba coupling is present in graphene-based heterostructures, the transition from WAL to  WL can be observed as a consequence of pseudospin-spin entanglement~\cite{mccann06,nestoklon14,wang22,sousa22,golub24}.

Importantly, even if the Rashba coupling vanishes on average, the random spatial fluctuations of Rashba field can exist and can strongly affect the transport properties of 2D systems~\cite{Glazov_PhysE}. For example, in semiconductor heterostructures, the inevitable fluctuations in the dopant ion density lead to the fluctuations of Rashba field~\cite{Sherman2003,Bindel_2016}, whereas in case of surface states of topological insulators, the structural defects can lead to the spatially random SOC and spin-momentum locking inhomogeneity~\cite{Dyrdal_PRL2020}.  In case of graphene and van der Waals heterostructures, the random Rashba spin-orbit coupling can originate from ripples of graphene sheet, adatoms and/or impurities adsorbed at the surface~\cite{Ertler2009,dugaev11,Zhang_2012}. Some inhomogeneity of the substrate and also mismatch between 2D crystals in heterostructures, can be a source of spatially fluctuating spin-orbit fields as well~\cite{Tuan2014}.

It has been shown that the random Rashba field affects spin relaxation and spin-dependent transport in 2D systems~\cite{Glazov_PhysE,Dugaev2009,dugaev11,Zhang_2012}. For example, the random Rashba field is responsible for the robust to impurity-scattering spin Hall effect~\cite{Dugaev2010,dyrdal12} and nonlinear anomalous Hall effect~\cite{dugaev12}.
Moreover, it has been shown that in semiconducting heterostructures with spatially fluctuating Rashba field, the spin-flip scattering processes can lead to a negative  magnetoresistance~\cite{dugaev12}. Importantly, this effect is purely classical, i.e., it can be observed beyond the weak localization regime. 

Negative magnetoresistance in graphene has been reported in theoretical and experimental studies, not only with regard to the WL experiments~\cite{mccann06,Hong_prb2011,Rein2015,Ando2019}. In fact, systems with randomly distributed adatoms~\cite{Hong_prb2011,Rein2015}, are good examples of systems, where the random spin-orbit field can appear and can be one of possible sources of the observed negative magnetoresistance.
Bearing in mind the experimental observations, 
we consider in this paper longitudinal electronic transport (i.e., transport along the external electric field) in a graphene monolayer with proximity-induced random Rashba spin-orbit interaction. Additionally, an external magnetic field is applied in the graphene plane and also parallel to the electric field.  Assuming the minimal model describing the low-energy electronic states near the Dirac points in graphene, we calculate the electron  momentum relaxation time  and the longitudinal conductivity and magnetoresistivity. To do this we use the Kubo formalism and Green function techniques and show, that  electron scattering on Rashba spin-orbit fluctuations leads to a negative magnetoresistivity, which is in agreement with experimental observations \cite{cite-key,PhysRevB.101.075425}.

\section{Model}

Bearing in mind the main objective of this paper, as described above in the Introduction, we
consider graphene in an external magnetic field applied in the graphene plane and in the presence of spatially fluctuating extrinsic Rashba spin-orbit interaction.    
The corresponding Hamiltonian describing  low-energy electron states in the vicinity of the Dirac point $K$  has the following form:
\begin{equation} 
\label{eq:H}
H=H_{0}+H_{\mathbf{B}}+H_R.
\end{equation} 
The term $H_{0}$, $H_{0} = v\, \btau \cdot {\mathbf{k}}$, describes pristine graphene with ${\mathbf{k}} = (k_{x}, k_{y})$ being 2D wavevector, while $\btau$ stands for the Pauli matrices acting  in the sublattice space. The second term, $H_{\mathbf{B}} = \bsig \cdot \bD$, is the  Zeeman term describing coupling with the external in-plane magnetic field. Here,  $\bsig$  are the Pauli matrices in the spin space, and  we assume  $\bD = (\Delta, 0, 0) $,
where $\Delta=g\mu_B B/2$  and the external magnetic field is along the axis $x$. Note, that by assuming the in-plane magnetic field, we avoid diamagnetic orbital effects. Therefore,  in Eq.~\eqref{eq:H} we take into account only the Zeeman coupling.

The last term, $H_R$, in Eq.~\eqref{eq:H} stands for the Rashba spin-orbit Hamiltonian,
\begin{equation} 
\label{1b}
H_R= \lambda({\bf r})\, (\sigma_y\tau_x-\sigma_x\tau_y),
%=\lambda ({\bf r})\left( \begin{array}{cc} 0 & -i\tau _- \\
%i\tau _+ & 0\end{array} \right) ,
\end{equation}
%$\tau _\pm =\tau _x\pm i\tau _y\, $, while
where $\lambda({\bf r})$ is the randomly fluctuating in space Rashba parameter, 
%describes the electron coupling to a spatially fluctuating random Rashba field  
with vanishing mean value, $\left\langle \lambda ({\bf
r})\right\rangle =0$, and the correlation function given by the formula~\cite{Glazov_PhysE}, 
\begin{equation} 
\label{4}
C_{\lambda\lambda }\left(\bf{r-r}^{\prime }\right) \equiv \langle
\lambda ({\bf r})\,\lambda ({\bf r'})\rangle
=\left\langle\lambda^{2}\right\rangle F({\bf r-r'}),
\end{equation}
where $F({\bf r-r'})$ is a certain correlation function. In the following, the Fourier transform of  $F({\bf r})$ will be  assumed in the  form $F(\mathbf{q})=\xi ^2\, e^{-\xi^{2} q^2}$, where $\xi $ is a characteristic correlation length.
%, and $\mathbf{q} = \mathbf{k} - \mathbf{k}'$. 

%\section{Unitary transformation} 

As the external magnetic field is oriented along the  axis $x$ (which is in the graphene plane), it is convenient to  perform a unitary transformation in the spin space, $H\to U^\dag HU$, with the transformation matrix defined as
\begin{eqnarray}
\label{eq:U}
U=\frac1{\sqrt 2}\left( \begin{array}{cc}
1 & i \\ 1 & -i \end{array}\right) ,
\end{eqnarray}
which transforms the spin operators in the following way: $U^\dag \sigma_xU= \sigma_z$ and $U^\dag \sigma_y U=-\sigma_x$.
Accordingly, we obtain  the transformed Hamiltonian in the following form:
\begin{eqnarray}
\label{7}
H_0 + H_{\mathbf{B}} 
= {\rm{diag}}(H_{0\uparrow },H_{0\downarrow }) ,
\label{eq:5}
\end{eqnarray}
where
\begin{eqnarray}
\label{eq:Hsig0}
H_{0\sigma }=v\, \btau \cdot {\bf k}+\Delta _\sigma ,  
\end{eqnarray}
with $\Delta _{\uparrow ,\downarrow }=\pm \Delta $, while the Rashba term upon  transformation acquires  the form,  
\begin{eqnarray}
\label{eq:HR}
H_R
= {\rm{diag}}\left(H^{\mathrm{(1)}}_{R\uparrow }, H^{\mathrm{(1)}}_{R\downarrow }\right) 
+ H^{\rm{(2)}}_{R},
%= H^{(1)}_{R\uparrow }\oplus H^{(1)}_{R\downarrow }+H^{(2)}_R ,
\end{eqnarray}
where
%$H^{{\rm{diag}}}_{R} = {\rm{diag}}\left(H^{\mathrm{diag}}_{R\uparrow }, H^{\mathrm{diag}}_{R\uparrow }\right)$ with 
\begin{eqnarray}
\label{eq:8}
&&H^{\mathrm{(1)}}_{R\uparrow ,\downarrow }=\mp \lambda ({\bf r})\, \tau _y ,
\end{eqnarray}
and
\begin{eqnarray}
&&H^{\mathrm{(2)}}_R=-\lambda ({\bf r})\, \sigma _x\tau _x .
%=-\lambda ({\bf r}) \left( \begin{array}{cc} 
%0 & \tau _x\\
%\tau _x & 0 \\ \end{array} \right) .
\label{eq:9}
\end{eqnarray}
Note that the states labeled here as $\uparrow $ ($\downarrow $) are in fact the states with spin oriented along (opposite to) the external magnetic field. 
%\ESc{Perhaps, it needs a better explanation since
%for the spin $x-$axis becomes $z-$axis after the transformation.}  

As follows from Eqs.~(\ref{eq:HR}), (\ref{eq:8}) and (\ref{eq:9}), after the unitary transformation one can clearly distinguish two components of the electron coupling with the fluctuating Rashba field. 
One of them 
%is described by $H^{(1)}_{R\uparrow , \downarrow }$ and 
corresponds to spin-conserving inter-sublattice scattering, and the other one %described by $H^{(2)}_R$ 
describes inter-sublattice spin-flip scattering processes. %These two types of scattering processes are not correlated.  
%to each other and can be described by random fields $\lambda _1({\bf r})$ and $\lambda _2({\bf r})$, respectively \textcolor{blue}{[AD: I think, we clarified that this is an exact form we do not need assume that there is no correlations between $\lambda_{1}$ and $\lambda_{2}$ as in further calculations there will be no cross terms $H_{R}^{{\mathrm{diag}}} H_{R}^{\mathrm{off}}$). Accordingly I would rather not introduce $\lambda_{1,2}$ and instead of them from beginning introduce only 'one' $\lambda$]}.  
In the following, the effects of random Rashba field are treated perturbatively as a source of electron scattering. To emphasize the role of scattering on Rashba spin-orbit fluctuations, we exclude other scattering processes, which however can be effectively included by a constant contribution to the imaginary part of the self-energy, if necessary.  

\begin{figure}[t]
\includegraphics[width=0.45\textwidth]{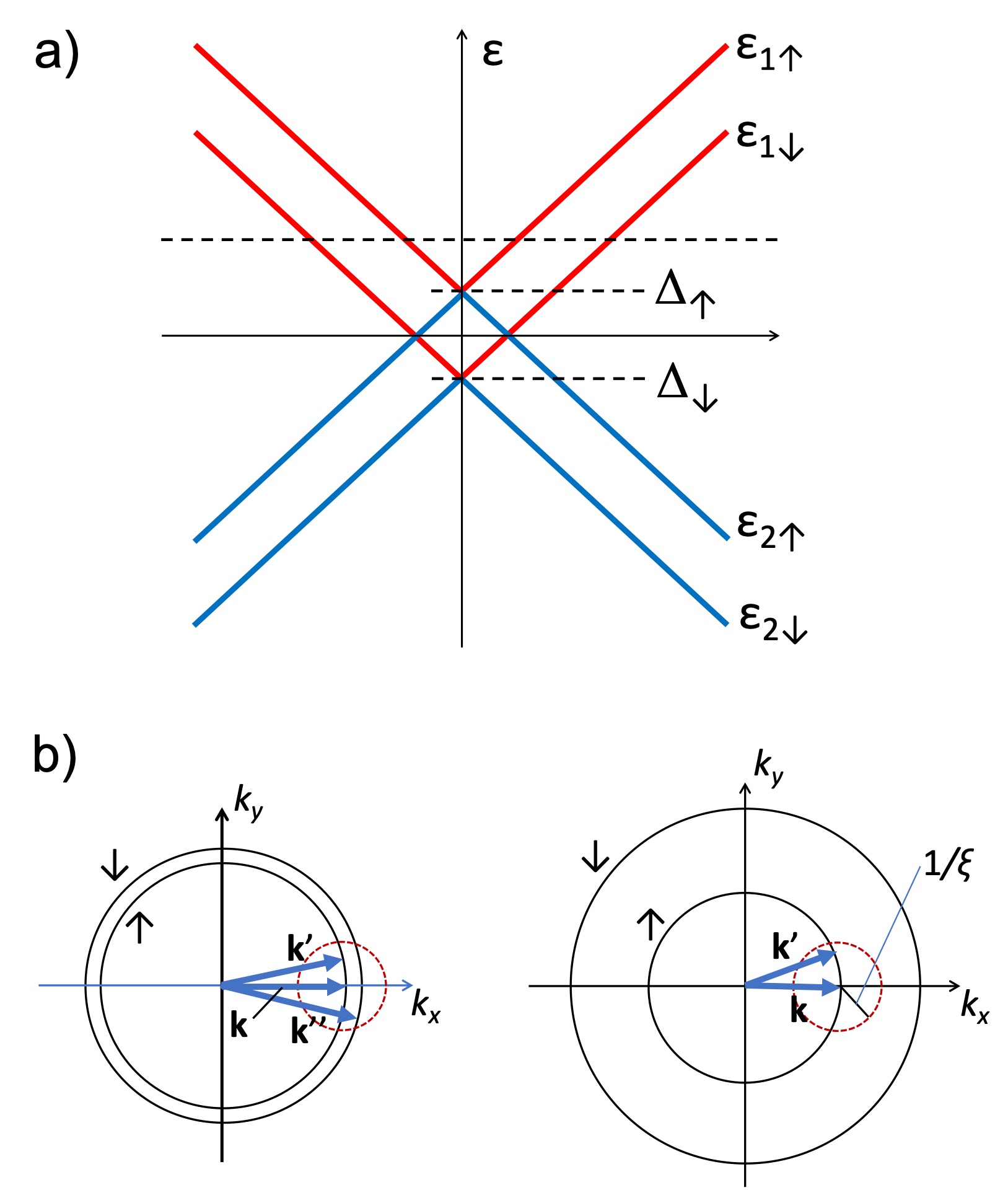}
\caption{(a) The electron energy spectrum of the transformed Hamiltonian. The Dirac cones are spin split and their two components are shifted up and down in energy scale. 
%\ESc{Suggest to add the top view and show schematically scattering processes, as in Fig. 1 in Ref. [57]. This will help a lot to understand the entire paper.}
(b) Schematic of possible scattering processes from the spin-up Fermi circle in weak (left) and strong (right) magnetic fields. The scatterings ${\bf k} \to {\bf k^{\prime}}$ (spin-conserved scattering) or ${\bf k} \to {\bf k^{\prime\prime}}$ (spin-flip scattering) are possible if $\Delta k<1/\xi $, where $\Delta k =|{\bf k^{\prime}} - {\bf k}| $  or $\Delta k =|{\bf k^{\prime\prime}} - {\bf k}| $, respectively.}
%here the difference in wavevectors $k$ associated with the two Fermi circles.
%\ESc{What is the definition (formula) of $\Delta k$ ?}}
\label{fig:Fig1}
\end{figure}

%Thus, after the unitary transformation we obtain the model of graphene in a perpendicular Zeeman field, with non-spin-flip and spin-flip random scatterers.  

The Green's function of the Hamiltonian $H_{0\sigma}$ ($\sigma =\uparrow , \downarrow$) has the following form:
\begin{equation}
\label{11}
G_{0\sigma }(\varepsilon ,{\bf k})=\frac{\varepsilon -\Delta _\sigma +\mu +v\, \btau \cdot {\bf k}}
{\prod\limits_{n=1,2} [\varepsilon - \varepsilon _{n\sigma }({\bf k})+\mu +i\delta \, {\rm sgn }\, \varepsilon ]},
%\nonumber \\ \times
%\frac1{(\varepsilon - \varepsilon _{2\sigma }(k)+\mu +i\delta \, {\rm sgn }\, \varepsilon )}\; ,
\end{equation}
where $\varepsilon _{1\sigma }({\bf k})=\varepsilon _{1\sigma }(k) =\Delta _\sigma +vk$ and 
$\varepsilon _{2\sigma }({\bf k})=\varepsilon _{2\sigma }(k)=\Delta _\sigma -vk$ are the eigen states of Hamiltonian (\ref{eq:5}) with $k=|{\bf k}|$,  while $\mu $ is the chemical potential. 
Note, that there is a $2\times 2$ unit matrix in front of the first three terms in the nominator, which according to our notation (also in the following) is not indicated explicitly. 
The energy spectrum is shown schematically in Fig.~\ref{fig:Fig1}. 
Linear dependence of the electron energy $\varepsilon $ on $k$ in graphene leads to the linear in $\varepsilon $ dependence of the density of states, $\rho (\varepsilon )\sim \varepsilon $. Accordingly, the magnetic-field-induced shift of spin subbands (like presented in Fig.~1) does not affect the total density of states $\rho _{\rm tot} =\rho _\uparrow +\rho _\downarrow $. 
The external magnetic field applied in the graphene plane shifts the chemical potential in order to conserve the particle number. When the chemical potential in the absence of magnetic field, denoted as $\mu_0$,   obeys  the condition  $\mu_0>\Delta$, then the  
chemical potential varies with  the field as $\mu=\sqrt{\mu_0^2-\Delta ^2}$. In turn, when $\mu_0<\Delta$, then the chemical potential is independent of the field, $\mu =\mu_0$.

\section{Relaxation processes due to random Rashba field}

To find the electron relaxation time, we need to calculate first the appropriate self-energy. Then, the relaxation times will be determined from the imaginary part of the self-energy.

\subsection{Self-energy}

\begin{figure}[t]
\includegraphics[width=0.49\textwidth]{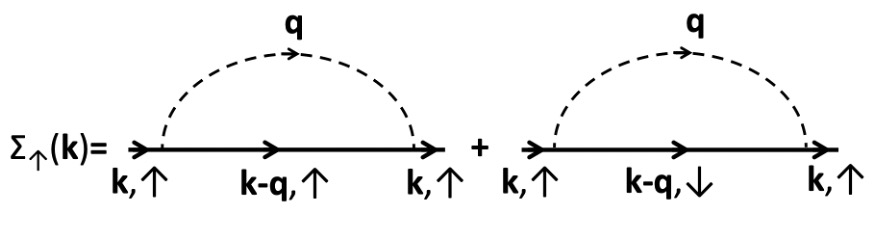}
\caption{Self energy due to the scattering from fluctuating Rashba coupling. The left (right) part corresponds to spin-conserving (spin-flip) scattering processes. }
\label{fig:Fig2}
\end{figure}

The self energy due to scattering of electrons from fluctuating spin-orbit interaction is given by the formula
\begin{equation}
\Sigma_{\mathbf{k}} = \int \frac{d^{2}{\bf k}'}{(2\pi)^{2}} H_{\mathbf{k} \mathbf{k}'}^{R} G_{0} (\varepsilon, \mathbf{k}') H_{\mathbf{k}' \mathbf{k}}^{R},
\end{equation}
where $H_{\mathbf{k} \mathbf{k}'}^{R}$ is the Bloch representation of the Hamiltonian~\eqref{eq:HR}.  Note, the self energy is here a $4\times 4$ matrix. Due to the form of Eq.~\eqref{eq:HR}, one  can factorize the self-energy as $\Sigma = {\rm{diag}}\left( \Sigma_\uparrow , \Sigma_\downarrow \right)$, and present  $\Sigma _\sigma$ as a sum of the contributions from spin-conserving and spin-flip scattering processes  (see also Fig.\ref{fig:Fig2}) in the following form:
\begin{eqnarray}
\label{12.1}
\Sigma _\sigma (\varepsilon ,{\bf k})
=\langle \lambda^2\rangle \int \frac{d^2{\bf k}'}{(2\pi)^2}\,
F(|{\bf k-k'}|) \tau_y G_{0\sigma}(\varepsilon, {\mathbf{k}'}) \tau_y 
\nonumber\\
+\langle \lambda^2\rangle \int \frac{d^2{\bf k}'}{(2\pi) ^2}\,
F(|{\bf k-k'}|)\, \tau _x  G_{0 \bar{\sigma}}(\varepsilon, \mathbf{k}') \tau_x,\,\, 
\end{eqnarray}
where we introduced the notation: $\bar{\sigma }= \downarrow$ when $\sigma = \uparrow$, and $\bar{\sigma }= \uparrow$ when $\sigma = \downarrow$.

The self energy $\Sigma _\sigma (\varepsilon ,{\bf k})$ is a $2\times 2$ matrix in the sublattice space, and its imaginary part determines the electron relaxation rates. Thus, the relaxation rate  of electrons with spin $\sigma$ is described by the $2\times 2$ matrix $\gamma _{\sigma }(\varepsilon , {\bf k_\sigma})$, which is the imaginary part of $\Sigma _\sigma (\varepsilon , {\bf k}_\sigma )$,
with $|{\bf k}_\sigma |=k_\sigma $ being the solution of the equation $\varepsilon _{n\sigma }(k)-\mu =\varepsilon $ for the two eigenstates $n=1,2$
In the following we neglect the real part of self energy as it leads to a small disorder-induced correction to the electron dispersion relations.   

Let us take $\mu > |\Delta_{\sigma}|$ and calculate $\gamma _{\sigma }(0,{\bf k_\sigma})$, corresponding to electrons at the Fermi level (see Fig.~\ref{fig:Fig1}). In this case, 
we find
\begin{eqnarray}
\label{16}
\gamma _{\sigma }(0,{\bf k_\sigma})
=\frac{\xi ^2\langle \lambda ^2\rangle}{4v}
\Big\{ e^{-2\xi ^2k_\sigma ^2}k_\sigma  
\Big[ I_0(2\xi ^2k_\sigma ^2) \nonumber \\
+\frac{\btau _1\cdot {\bf k}_\sigma }{k_\sigma }\, I_1(2\xi ^2k_\sigma ^2) \Big] 
%\nonumber \\
+e^{-\xi ^2(k_\sigma ^2+k_{\bar{\sigma }}^2)}\, k_{\bar{\sigma }}  
\Big[ I_0(2\xi ^2k_\sigma k_{\bar{\sigma }}) \nonumber \\
-\frac{\btau _1\cdot {\bf k}_\sigma }{k_\sigma }\, 
I_1(2\xi ^2k_\sigma k_{\bar{\sigma }})\Big] \Big\},\hspace{0.5cm} 
\end{eqnarray}
%where we used the notations
with $k_\sigma  =\mu _\sigma /v$, $\mu _\sigma =\mu -\Delta _\sigma $, and $\btau _1=(-\tau _x, \tau _y)$.
%, and $\btau _2=(\tau _x, -\tau _y)=-\btau _1$. 
%\ESc{$\btau _2$ is not used here.} 
%
Apart from this,  
$I_0(z)$ and $I_1(z)$ in Eq.~(\ref{16}) are the modified Bessel functions, 
$$I_0(z)=\frac1{\pi }\int _0^\pi e^{z\cos \theta }d\theta ,\; $$    
$$I_1(z)=\frac1{\pi }\int _0^\pi e^{z\cos \theta }\cos \theta \, d\theta .$$

\subsection{Momentum relaxation time } 

The imaginary part of self energy is a matrix.
To find the relaxation time of electron in a certain state $|n\sigma {\bf k}\rangle $, we should solve the eigenvalue equation, which takes into account the self energy part,
\begin{eqnarray}
\label{14}
\det \big[ H_{0\sigma }({\bf k})-\mu +i\gamma _\sigma (\varepsilon , {\bf k}_\sigma )-\varepsilon \big] =0.
\end{eqnarray}
From now the dependence of $\gamma _\sigma$ on ${\bf k}$ is not indicated explicitly and this dependence will be restored when necessary.
Since we assume that the effect of $\gamma _\sigma $ is relatively weak, we can consider it as a small correction to the solutions $\varepsilon =\varepsilon _{1\sigma }(k)-\mu $ and $\varepsilon =\varepsilon _{2\sigma }(k)-\mu $ (i.e.,  the solutions of Eq.~(\Ref{14}) for $\gamma _\sigma =0$). Correspondingly, from Eq.~(\Ref{14}) we obtain the following equation:
\begin{eqnarray}
\det \left[
\Delta _\sigma +v\btau \cdot {\bf k}-\mu
+i\gamma _\sigma \big( \varepsilon _{1,2\sigma }-\mu )-\varepsilon \right]	=0,
\end{eqnarray}
where we choose $ \gamma _\sigma (\varepsilon) = \gamma _\sigma \big( \varepsilon _{1\sigma }-\mu )$ for the solution $\varepsilon \simeq \varepsilon _{1\sigma }(k)-\mu $ or $\gamma _\sigma (\varepsilon) = \gamma _\sigma \big( \varepsilon _{2\sigma }-\mu )$ for the second solution $\varepsilon \simeq \varepsilon _{2\sigma }(k)-\mu $. 

In the first case (i.e., for $n=1$), we take $\varepsilon =+0$ (which corresponds to electron states at the Fermi level, see Fig.~\ref{fig:Fig1}) and use $\gamma _\sigma (0)$ from Eq.~(13). Then we get
\begin{eqnarray}
\label{eq:21}
\varepsilon_{1\sigma }(k)=\Delta _\sigma + vk_\sigma +iw'_\sigma -iw''_\sigma \, \frac{k_{x\sigma }^2-k_{y\sigma }^2}{k_\sigma } \, ,
\end{eqnarray}
where we introduced the following notation:  
\begin{eqnarray}
\label{19}	
&& w'_\sigma =\frac{\xi ^2\langle \lambda ^2\rangle}{4v}
\Big\{ e^{-2\xi ^2k_\sigma ^2}k_\sigma  
I_0(2\xi ^2k_\sigma ^2) \nonumber \\
&&\hspace{2.5cm}
+e^{-\xi ^2(k_\sigma ^2+k_{\bar{\sigma }}^2)}\, k_{\bar{\sigma }}  
I_0(2\xi ^2k_\sigma k_{\bar{\sigma }}) \Big\} ,
\\
&& w''_\sigma =\frac{\xi ^2\langle \lambda ^2\rangle}{4vk_\sigma }
\Big\{ e^{-2\xi ^2k_\sigma ^2}\, k_\sigma \,
I_1(2\xi ^2k_\sigma ^2) 
\nonumber \\
&&\hspace{2.5cm}
-e^{-\xi ^2(k_\sigma ^2+k_{\bar{\sigma }}^2)}\, k_{\bar{\sigma }}\, 
I_1(2\xi ^2k_\sigma k_{\bar{\sigma }})\Big\} .
\end{eqnarray}
From Eq.~\eqref{eq:21} we find the following expression for the anisotropic relaxation time $\tilde{\tau }$,
\begin{eqnarray}
\label{22}
\frac{\hbar }{2\tilde{\tau }_{1\sigma }({\bf k}_\sigma )}\equiv \gamma _{1\sigma }(0,{\bf k}_\sigma )=w'_\sigma -w''_\sigma \, \frac{k_{x\sigma }^2-k_{y\sigma }^2}{k_\sigma } \, .
\end{eqnarray}

Similarly, one can find the relaxation time of a hole with energy $\varepsilon =-2\mu $. For this purpose we have to use Eq.~(\ref{12.1}) and calculate $\gamma _\sigma (-2\mu )$. As a result we find $\gamma _\sigma (-2\mu )=-w'_\sigma +w''_\sigma \, (k^2_{x\sigma }-k^2_{y\sigma })/k_\sigma $. 
Taking into account that 
$\hbar /2\tau _{2\bar{\sigma }}=-\gamma _{2\bar{\sigma }}(-2\mu )$ we come to equality $\tilde{\tau}_{2\bar{\sigma }}({\bf k})
=\tilde{\tau}_{1\sigma }({\bf k})$, as one might expect from the electron-hole symmetry. 

Importantly, the anisotropy of relaxation time is related to the orientation of external magnetic field. 
When $k_{\bar{\sigma }}\to k_\sigma $ we get $w^{\prime\prime}_\sigma \to 0$, and the relaxation anisotropy disappears. Specifically, for $\mu \gg \Delta _\sigma $ 
(i.e., when the magnetic field is relatively weak) one can neglect this anisotropy, taking $\gamma _{1,2\sigma }({\bf k}_\sigma )=w'_\sigma $, and the electron relaxation time related to the scattering from fluctuating Rashba SO interaction is $\tilde{\tau }_\sigma (k_\sigma )\simeq \hbar/2w'_\sigma $. In the following we analyze  the $\tilde{\tau}_{1\sigma }({\bf k})$ and suppress the index 1. 

\begin{figure}[t]
\includegraphics[width=0.47\textwidth]{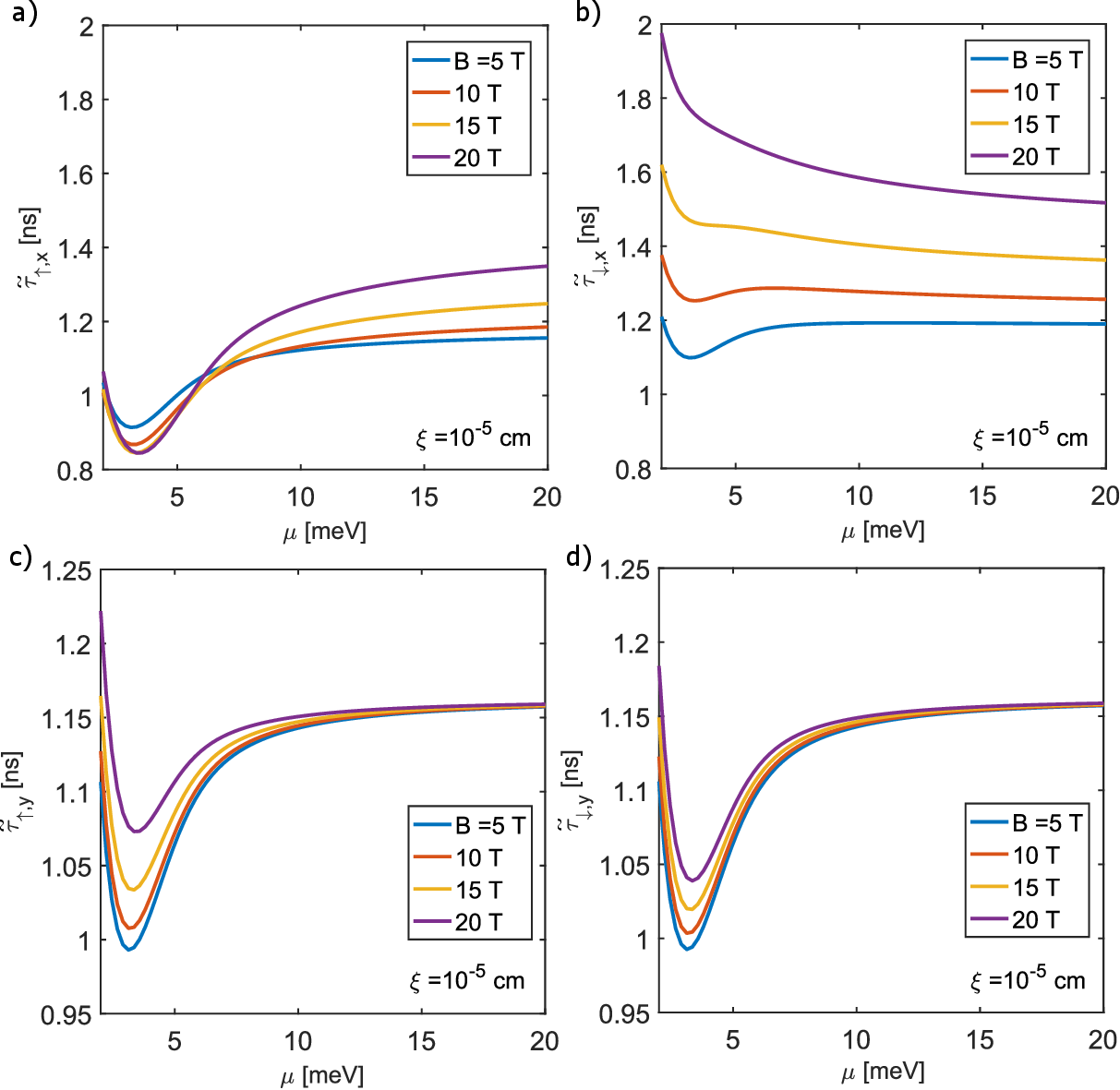}
\caption{(a,b) Relaxation time of an $\uparrow (\downarrow)$-electron in the state corresponding to the wavevector ${\bf k}_{\uparrow}$ (${\bf k}_{\downarrow })$ oriented along the magnetic field, presented  as a function of the chemical potential $\mu$ for indicated values of magnetic field. (c,d) Relaxation time of an $\uparrow (\downarrow)$-electron in the state corresponding to the wavevector ${\bf k}_{\uparrow}$ (${\bf k}_{\downarrow })$ oriented perpendicularly to the magnetic field, presented as a function of the chemical potential $\mu$ for indicated values of magnetic field.  The other parameters in (a-d) are:  $v=5\times 10^{-8}$~eV cm, $\langle \lambda ^2\rangle^{1/2} =0.1$~meV, and $\xi$ as indicated. 
}
\label{fig:Fig3}
\end{figure}

In Fig.~\ref{fig:Fig3} we present the relaxation time as a function of the chemical potential for electrons in the spin-$\uparrow$ and spin-$\downarrow$ subbands, and for the wavevectors along the magnetic field ($\tilde{\tau}_x$, top panels) and normal to the magnetic field ($\tilde{\tau}_y$, bottom panels). Different curves correspond to the indicated values of the magnetic field $B$, while the correlation length $\xi$ is fixed, $\xi=10^{-5}$ cm. 

Consider first the numerical data on the relaxation time in the spin-$\uparrow$ subband [subband with the smaller Fermi circle, see Fig.~\ref{fig:Fig2}(b)], and for the wavevectors along the magnetic field, $\tilde{\tau}_{\uparrow x}$, shown in Fig.~\ref{fig:Fig3}(a). Let us consider first the data for small values of $B$, say $B=5$ T in Fig.~\ref{fig:Fig3}(a). In the range of small chemical potential $\mu$, the relaxation times drop down with increasing $\mu$, reach minima around $\mu\approx 3$ meV, and then increase with a tendency to saturate for $B>10$ T. To understand  this behavior, we note that the range of momentum transfer during the scattering events is roughly determined by the circle of radius $1/\xi$, see Fig.~\ref{fig:Fig1}. This holds for scattering of electrons in both spin subbands (both Fermi circles). We also recall that the description is applicable for $\mu \ge \Delta$, where the two Fermi circles for both spin subbands are shown schematically in Fig.~\ref{fig:Fig1}. Moreover, for the assumed parameters, $1/\xi$ is comparable to Fermi wavevector for $\mu\approx 10$ meV (for $B=0$). Thus, in the limit of small chemical potential, strictly when $\mu = \Delta$, the Fermi circle for $\uparrow$-spin subband reduces to a point and there is no spin-flip scattering in the system. When the chemical potential is slightly larger than $\Delta$, $\mu \ge \Delta$, the  circle of radius $1/\xi$ associated with the larger Fermi circle (spin-$\downarrow$ electrons) covers the small Fermi circle. Similarly, the circle of radius  $1/\xi$ associated with the smaller Fermi circle (spin-$\uparrow$ electrons) covers the larger Fermi circle, or its significant part. Accordingly, the spin-flip scattering processes are admitted for $\mu \ge \Delta$. 
Consequently, the relaxation time of spin-$\uparrow$  decreases initially with increasing $\mu$ due to the presence of spin-flip scattering and its contribution to the electron scattering rate. The spin-flip scattering rates  increase with increasing $\mu$, so the the corresponding relaxation time decreases initially with increasing  $\mu$. With a further increase in $\mu$, the number of states available for backward spin-flip and spin-conserved  scattering becomes reduced, while the forward scattering processes survive due to small magnetic field (small separation of the Fermi circles). Thus upon reaching a minimum, the relaxation time increases again, with a tendency to  saturate with increasing $\mu$. 

When the magnetic field $B$  is increased, the separation of the two Fermi circles increases as well, so the radius of the smaller Fermi circle becomes decreased for a constant $\mu$, see Fig.~\ref{fig:Fig2}(b).
Accordingly, the number of electron states to which a spin-$\uparrow$ electron can be scattered is reduced  in comparison with that at $B=0$, and one observes a small shift of the minimum in the relaxation time towards larger $\mu$  and also as small reduction of the relaxation time in the minimum with increasing the field $B$, see Fig.~\ref{fig:Fig3}(a) for small values of $\mu$. For significantly larger values of $\mu$, the spin-flip scattering as well as the backward scattering become reduced  with increasing $B$, so the relaxation time increases with $\mu$ and also with $B$, as shown in Fig.~\ref{fig:Fig3}(a).

Behaviour of the relaxation times for spin-$\downarrow$ electrons with wavevector along the magnetic field, see Fig.~\ref{fig:Fig3}(b), as well as for the  spin-$\uparrow$ and spin-$\downarrow$ electrons with wavevectors normal to the magnetic field are shown,   Fig.~\ref{fig:Fig3}(c) and Fig.~\ref{fig:Fig3}(d), respectively, can be accounted for in a similar way. General features are qualitatively similar to those shown in Fig.~\ref{fig:Fig3}(a) and described above. However,  some differences follow from the geometry of the system and scattering asymmetry, see Eq.~\ref{22}. For instance, the minimum in the relaxation time shown in  Fig.~\ref{fig:Fig3}(b) is less pronounced tan that in Fig.~\ref{fig:Fig3}(a), and disappears at large values of $B$. Note, the difference between relaxation times for spin-$\uparrow$ and spin-$\downarrow$ with wavevectors normal to the magnetic field is less pronounced than in the case of wavevectors along the magnetic field.

\begin{figure}[t]
%\vskip-0.5cm
\includegraphics[width=0.47\textwidth]{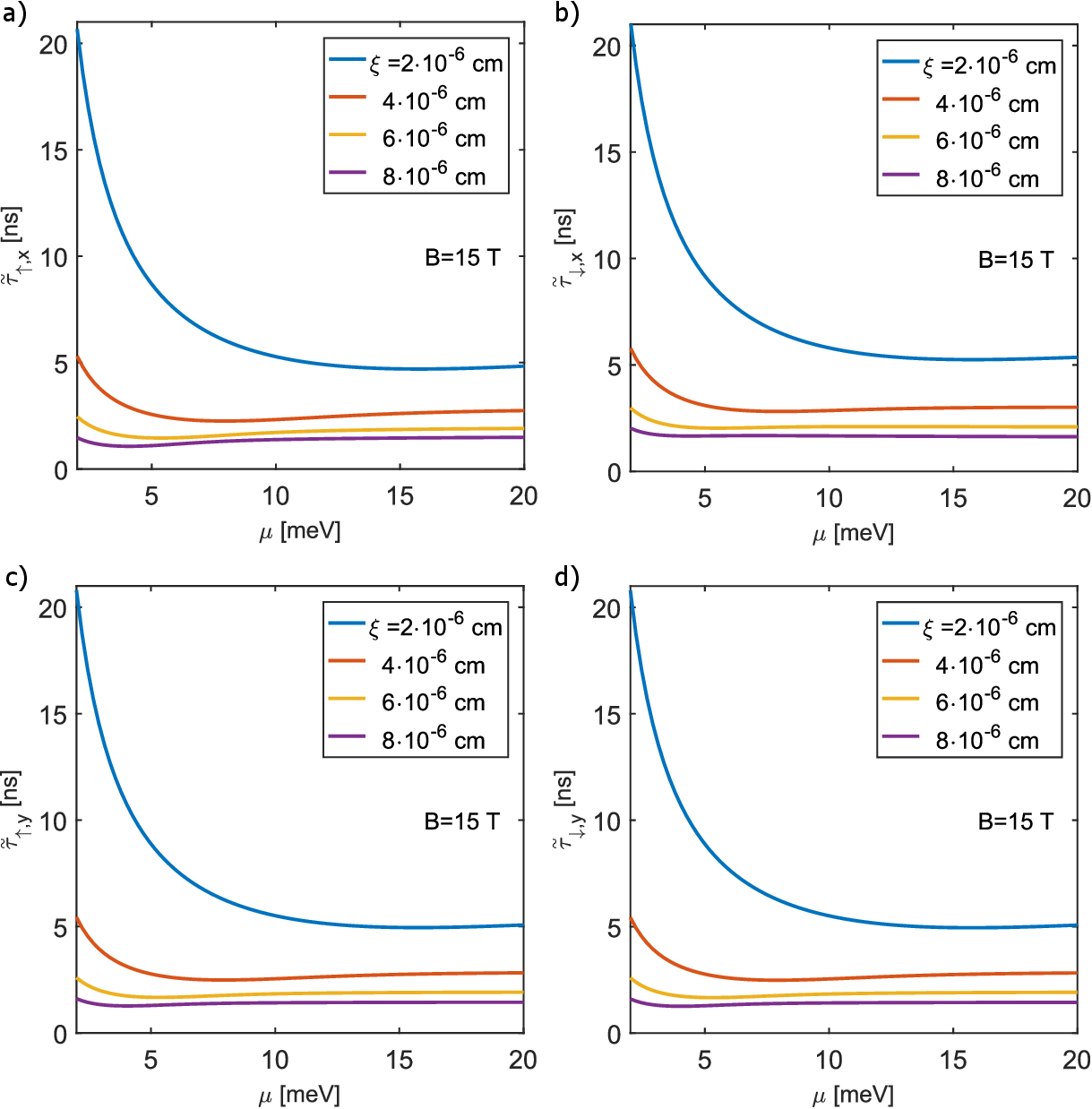}
\caption{
(a,b) Relaxation time of an $\uparrow (\downarrow)$-electron in the state corresponding to the wavevector ${\bf k}_{\uparrow}$ (${\bf k}_{\downarrow })$ oriented along the magnetic field, presented  as a function of the chemical potential $\mu$ for indicated values of $\xi$. 
(c,d) Relaxation time of an $\uparrow (\downarrow)$-electron in the state corresponding to the wavevector ${\bf k}_{\uparrow}$ (${\bf k}_{\downarrow })$ oriented perpendicularly to the magnetic field, presented as a function of the chemical potential $\mu$ for indicated values of $\xi$.
The other parameters in (a-d):  $v=5\times 10^{-8}$~eV cm, $\langle \lambda ^2\rangle^{1/2} =0.1$~meV, and $B$ as indicated. 
}
\label{fig:Fig4}
\end{figure}

In Fig.~\ref{fig:Fig4} we show the relaxation times for the wavevector along magnetic field, $\tilde{\tau}_{\sigma x}$, ($\sigma = \uparrow , \downarrow$) and perpendicular to the magnetic field, $\tilde{\tau}_{\sigma y}$, as a function of the chemical potential for  the  fixed magnetic field, $B=15$ T. Different curves correspond  to the indicated values of the correlation length $\xi$, which are taken smaller than that in Fig.~\ref{fig:Fig3}, more specifically $10^{-5}\mbox{ cm} >\xi > 10^{-6}\mbox{ cm}$.  Since scattering is proportional to $\xi^2$, see the correlation function below Eq.~(3), the relaxation times in Fig.~\ref{fig:Fig4} are significantly longer than those in Fig.~\ref{fig:Fig3}. Moreover, the shape of all the curves shown in Fig.~\ref{fig:Fig4} are qualitatively similar. This follows from the fact, that for $\xi =10^{-6}$ cm, $1/\xi$ is much larger than the Fermi  wavevectors, and thus all states are available for scattering processes. Therefore, the effects of specific electronic subband structure of graphene on the relaxation times become washed out and the system behaves roughly as a conventional 2D electron gas.   

From the above results follows, that the relaxation times generally increase with increasing magnetic field (see  Fig.~\ref{fig:Fig3}). Some deviations from this behaviour occur only for $\uparrow$-electrons in the small chemical potential regime, see  Fig.~\ref{fig:Fig3}(a). In turn, the relaxation times decrease with increasing correlation length $\xi$  (see Fig.~\ref{fig:Fig4}), i.e., with increasing strength of scattering on spin-orbit disorder.
From this behavior one may expect that the conductivity increases with increasing magnetic field and decreases with increasing spin-orbit disorder. From this, in turn, one may conclude that the corresponding magnetoresistance is negative. 
To show this explicitly, we calculate now the longitudinal conductivity, $\sigma_{xx}$.     

\section{Electronic transport and magnetoresistance }

%\ESc{Is it worth mentioning transport scattering time also ?}

\begin{figure}[t]
\includegraphics[width=0.475\textwidth]{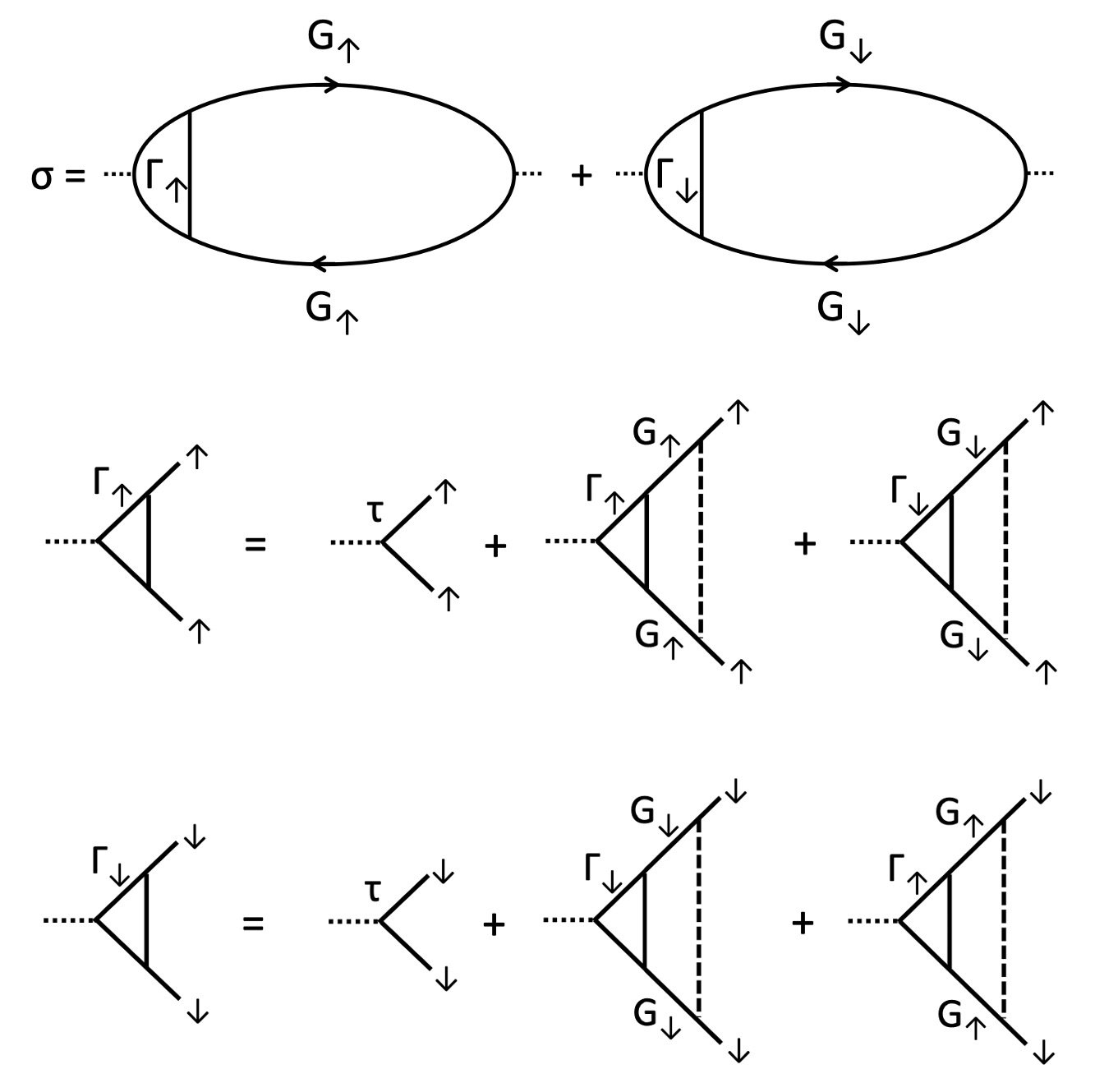}
\caption{Kubo diagrams for the conductivity and diagrams representing equations for the vertex functions. }
\label{fig:Fig5}
\end{figure}

The charge current density,  as a response of the system to the electromagnetic field described by the vector potential ${\bf A}(t)={\bf A}_\omega e^{-i\omega t}$,
can be written (due to the gauge transformation ${\bf k}\to {\bf k}-e{\bf A}/\hbar c$)  in the linear response formalism as
\begin{eqnarray}
\label{25}
j_x(\omega )= - i\sum_\sigma {\rm Tr}
  \int \frac{d^2{\bf k}}{(2\pi )^2}\int\frac{d\varepsilon }{2\pi }\,
J_{\mathbf{k} \sigma x}(\varepsilon, \omega )
\nonumber \\ \times
G_{\mathbf{k} \sigma }(\varepsilon +\hbar \omega )\, H^{\mathbf{A}}_{\sigma}\, G_{\mathbf{k} \sigma }(\varepsilon ),
\end{eqnarray}
where 
\begin{eqnarray}
\label{24}
H^{\bf{A}}_{\sigma}
%=-\frac{ev\, \btau \cdot {\bf A}}{\hbar c} \, 
= - \mathbf{j}_{\sigma}\cdot\mathbf{A}_{\omega}
\end{eqnarray}
describes  coupling of the system to an external electromagnetic field, ${\bf A}_\omega ={\bf E}_\omega /i\omega $, within the minimal coupling approach,
the charge current density, $\mathbf{j}_\sigma$, is  defined as 
\begin{equation}
j_{\sigma i} = e v_{\sigma i}=\frac{e}{\hbar}\, \frac{\partial H_{0 \sigma}}{\partial k_i}=\frac{ev}{\hbar} \tau_i ,
\end{equation}
whereas $J_{\mathbf{k} \sigma x}(\varepsilon, \omega )$ is the renormalized velocity vertex function (see Fig.~\ref{fig:Fig5}~(a)) defined as
\begin{equation}
\label{eq:Jvertex}
J_{\mathbf{k} \sigma x}(\varepsilon, \omega ) = \frac{ev}{\hbar} \Gamma_{\mathbf{k} \sigma x}(\varepsilon, \omega ) ,
\end{equation}
with the vertex function $\Gamma_{\mathbf{k} \sigma x}$ given below. 

%\subsection{Conductivity}
We assume the electric field along the axis $x$, i.e., parallel to the magnetic field, and focus on the longitudinal conductivity.  
Taking into account the relation $j_{x}(\omega) = \sigma_{xx} E_{\omega x}$, one finds 
\begin{eqnarray}
\label{29}
\sigma _{xx}(\omega )
=\frac{e^2v^2}{\hbar ^2\omega }\, {\rm Tr}
\sum _\sigma \int \frac{d^2{\bf k}}{(2\pi )^2}\frac{d\varepsilon }{2\pi }\,
\Gamma _{{\bf k}\sigma x}(\varepsilon, \omega ) 
\nonumber \\ \times
G_{\bf k\sigma }(\varepsilon +\hbar \omega )\, \tau _x\, 
G_{\bf k\sigma }(\varepsilon ).
\end{eqnarray}
The vertex function $\Gamma _{{\bf k}\sigma x}(\varepsilon,\omega ) $ is presented in Fig.~\ref{fig:Fig5} and takes the following form:
\begin{eqnarray}
\label{26}
\Gamma _{{\bf k}\sigma x}(\varepsilon,\omega ) 
=\tau _x+\langle \lambda ^2\rangle \int \frac{d^2{\bf k}'}{(2\pi )^2}\, F(|{\bf k-k'}|)\, 
\tau _y\, G_{{\bf k}'\sigma }(\varepsilon ) 
\nonumber \\ \times
\Gamma _{{\bf k}'\sigma x}(\varepsilon ,\omega )\, 
G_{{\bf k}'\sigma }(\varepsilon +\hbar \omega )\, \tau _y  
\nonumber \\
+\langle \lambda ^2\rangle \int \frac{d^2{\bf k}'}{(2\pi )^2}\, F(|{\bf k-k'}|)\, \tau _x
G_{{\bf k}'\bar{\sigma }}(\varepsilon )\, 
\nonumber \\ \times
\Gamma _{{\bf k}'\bar{\sigma }x}(\varepsilon ,\omega )\, 
G_{{\bf k}'\bar{\sigma }}(\varepsilon +\hbar \omega )\, \tau _x\, . \hskip0.5cm	
\end{eqnarray}
In the limit of $\omega \to 0$, the main contribution to the current is from the Fermi surface, $\varepsilon \to 0$. In this limit, the equation for the vertex function reads
\begin{eqnarray}
\label{27}
\Gamma _{{\bf k}\sigma x} 
=\tau _x+\langle \lambda ^2\rangle \int \frac{d^2{\bf k}'}{(2\pi )^2} F(|{\bf k-k'}|)\, \tau _y
G^A_{{\bf k}'\sigma } \Gamma _{{\bf k}'\sigma x} G^R_{{\bf k}'\sigma } \tau _y 
\nonumber \\
+\langle \lambda ^2\rangle \int \frac{d^2{\bf k}'}{(2\pi )^2}\, F(|{\bf k-k'}|)\, \tau _x 
G^A_{{\bf k}'\bar{\sigma }}\, \Gamma _{{\bf k}'\bar{\sigma }x}\, G^R_{{\bf k}'\bar{\sigma }} \, \tau_x , \hskip0.5cm	
\end{eqnarray}  
where we used the notation, $\Gamma _{{\bf k}\sigma x}\equiv \Gamma _{{\bf k}\sigma x}(0,0)$ and $G^{R,A}_{{\bf k}\sigma }\equiv G^{R,A}_{{\bf k}\sigma }(0)$. The derivation of the vertex function is presented in Appendix~\ref{App:A}. 

As a result, in the limit of $\omega \to 0$ we find the following expression for the static (dc)  conductivity: 
\begin{eqnarray}
\label{30}
\sigma _{xx}=\frac{e^2v^2}{2\pi \hbar }\, {\rm Tr}
\sum _\sigma \int \frac{d^2{\bf k}}{(2\pi )^2}\,
\Gamma _{{\bf k}\sigma x}\, G^R_{\bf k\sigma }\, \tau _x\, G^A_{\bf k\sigma }\, ,
\end{eqnarray}
which includes the vertex function $\Gamma _{{\bf k}\sigma x}$.
To solve the self-consistent vertex equations we postulate the solution in the following form:
\begin{eqnarray}
\label{eq:a2}
\Gamma _{{\bf k}\sigma x}=\tau _x\, g_{1\sigma {\bf k}}+\tau _y\, g_{2\sigma {\bf k}}+g_{3\sigma {\bf k}}\, ,	
\end{eqnarray} 
where $g_{i\sigma {\bf k}}$ are certain functions of ${\bf k}$. The explicit forms of $g_{i\sigma {\bf k}}$ are given in the Appendix~\ref{App:A}.

For $\mu$ significantly larger than $\Delta_\sigma$, when the anisotropy of relaxation time is negligible and $\gamma_\sigma$  becomes constant, we find
\begin{eqnarray}
\label{42}
\sigma _{xx}
=\frac{e^2v^2}{\pi \hbar } \sum _\sigma \int \frac{d^2{\bf k}}{(2\pi)^2}\,\mathcal{D}_{\sigma}\left[ (\mu _\sigma ^2+v^2k_x^2-v^2k_y^2) g_{1 \sigma {\bf k}} \right. \nonumber\\
\left. + 2v^2k_xk_yg_{2\sigma {\bf k}} + 2\mu _\sigma vk_xg_{3\sigma {\bf k}}\right], \hspace{0.5cm} 
\end{eqnarray}
where
\begin{eqnarray}
\mathcal{D}_{\sigma} = \left[(\mu_\sigma -vk+i\gamma_\sigma )(\mu_\sigma +vk+i\gamma_\sigma )\right. \hspace{2.0cm} \nonumber\\
 \times
\left.(\mu_\sigma -vk-i\gamma_\sigma )(\mu_\sigma +vk-i\gamma_\sigma )\right]^{-1}.\hspace{0.5cm}
\end{eqnarray}
Then we get 
\begin{eqnarray}
\label{43}
\sigma _{xx}
=\frac{e^2}{8\pi ^2\hbar } \sum _\sigma \frac{\mu _\sigma }{\gamma _{1\sigma }}
\int _0^{2\pi }\; d\phi \; 
\big( \cos ^2\phi \, g_{1\sigma k_\sigma }
\nonumber \\
+\sin \phi \cos \phi \, g_{2\sigma k_\sigma }+\cos \phi \, g_{3\sigma k_\sigma }\big) ,
\end{eqnarray}
where we denoted $k_x=k\cos \phi \; $, $k_y=k\sin \phi$, so that $g_{n\sigma k}$ are functions of angle $\phi$. 
When the random Rashba field is small, $g_{1\sigma k}\to 1$, $g_{2\sigma k}=g_{3\sigma k}\to 0$, and for $\mu$ significantly larger than $\Delta_\sigma$, we finally get
\begin{eqnarray}
\label{44}
\sigma _{xx}^{0}=\frac{e^2}{8\pi \hbar } \sum _\sigma \frac{\mu -\Delta _\sigma }{\gamma _{\sigma }+\gamma_0}\; ,
\end{eqnarray}
where to avoid singularity when $\gamma_\sigma\to 0$),  we have added in the denominator  a constant term $\gamma_0$, which takes into account scattering by other defects. Such scattering becomes dominant when Rashba spin-orbit fluctuations disappear. However, in the following numerical calculations  we assumed the parameters of Rashba field fluctuations large enough, to make the scattering on spin-orbit fluctuations dominant, and thus we neglect other scattering processes.  

The effect of magnetic field is usually described quantitatively by the relevant magnetoresistance (MR), which  is defined by the standard formula,
\begin{equation}
{\rm{MR}}(B) = \frac{\rho_{xx}(B) - \rho_{xx}(0)}{\rho_{xx}(0)} \simeq \frac{\sigma_{xx}(0)}{\sigma_{xx}(B)} - 1.
\end{equation}
The magnetoresistance in conventional systems (with scalar scattering potential of impurities)  is usually positive. Negative magnetoresistance appears in some specific cases, like due to  weak localization or due to spin disorder, or in Kondo systems. 
Here we show that negative  magnetoresistance also appears due to scattering on fluctuating in space Rashba interaction.   

\begin{figure}[t]
%\vskip-0.5cm
\includegraphics[width=0.47\textwidth]{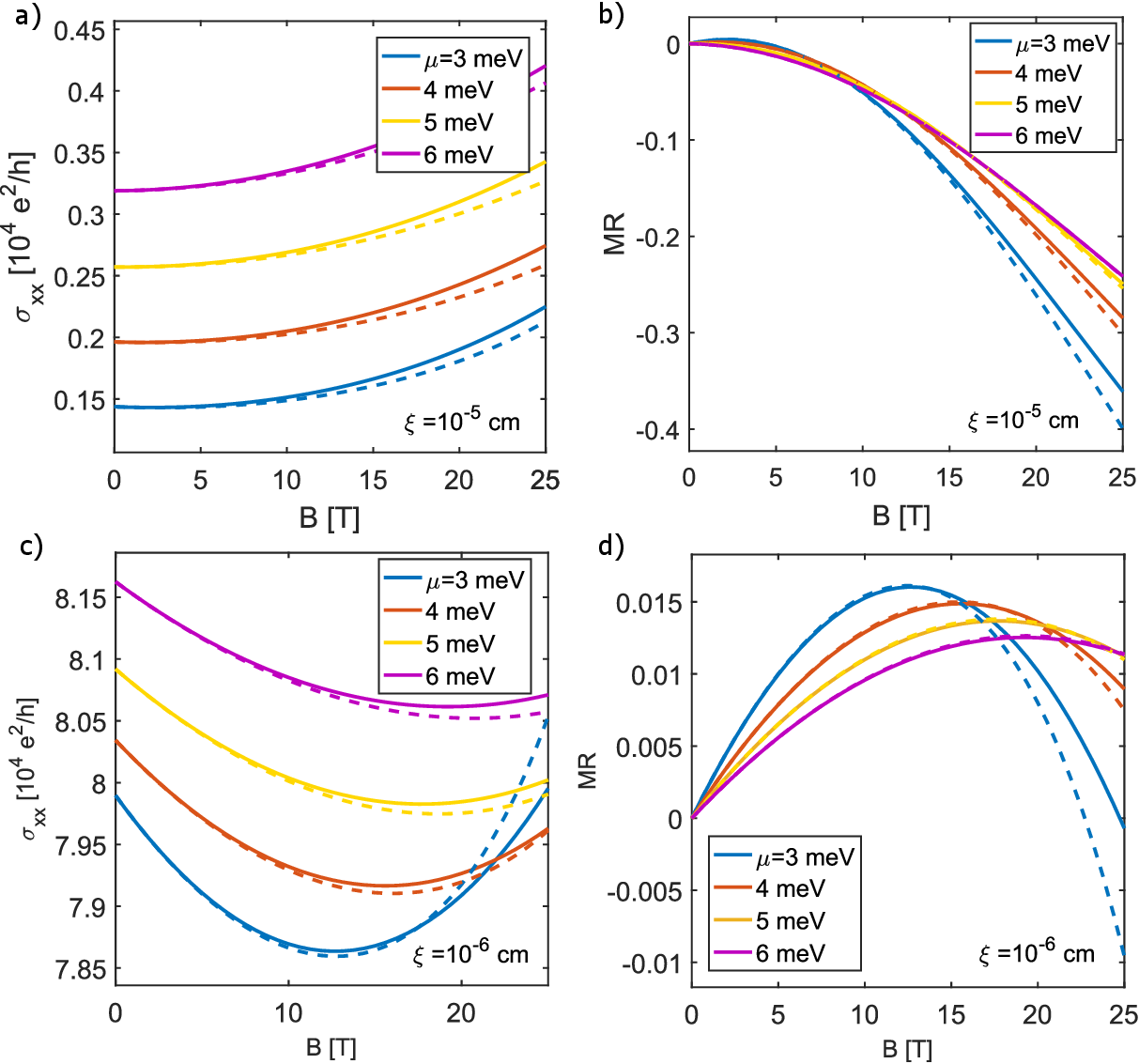}
\caption{Dependence of the conductivity  
and magnetoresistance on the magnetic field $B$ for indicated values of the  chemical potential $\mu $. Both Dirac points are here included. We use the parameters:  $v=5\times\,10^{-8}$~eV cm, $\langle \lambda ^2\rangle^{1/2} =0.1$~meV, while
$\xi = 10^{-5}$ cm  (a,b) and $\xi = 10^{-6}$ cm (c,d).  The solid lines correspond to the case of constant chemical potentials while the dashed lines correspond to the limit of conserved particle number.   
}
\label{fig:Fig6}
\end{figure}

\begin{figure}[t]
\includegraphics[width=0.48\textwidth]{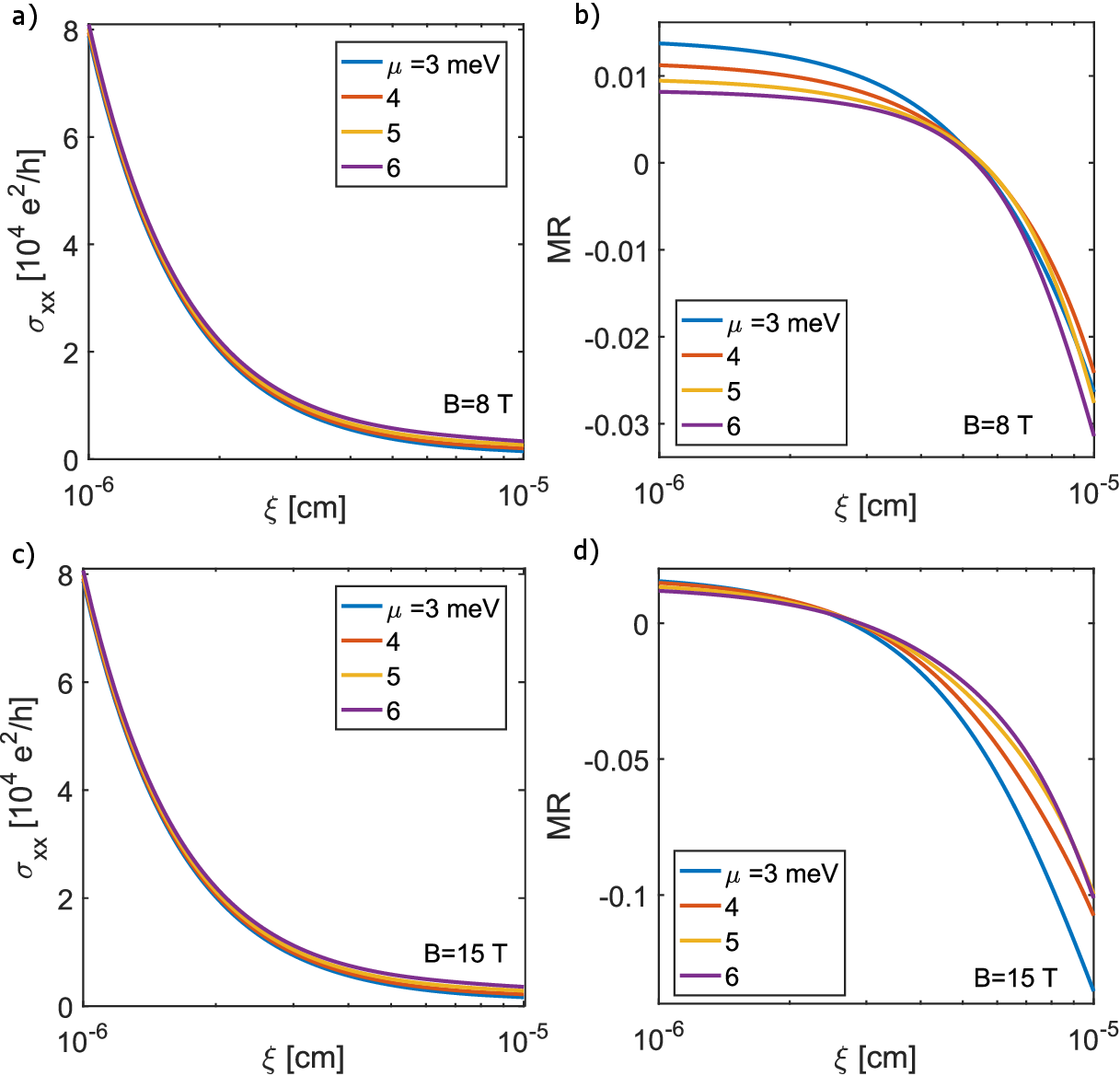}
\caption{Dependence of the longitudinal conductivity and magnetoresistance on the correlation length $\xi$ for two different values of the magnetic field, and chemical potential $\mu $ as indicated. Other parameters as in Fig.~\ref{fig:Fig6}. 
}
\label{fig:Fig7}
\end{figure}

Numerical results on the longitudinal conductivity and magnetoresistance in the presence of Rashba field fluctuations are shown in Fig.~\ref{fig:Fig6} as a function of the external magnetic field, and for two different values of the correlation length $\xi$. The solid lines correspond there to constant values of chemical potential $\mu =\mu_0$, while the dashed lines to the chemical potentials varying with  the field as $\mu=\sqrt{\mu_0^2-\Delta ^2}$. This formula holds for $\mu_0>\Delta$ considered in this paper. The latter case (dashed lines) corresponds to conserved particle number, while the former case (solid lines) corresponds to constant chemical potential (particle number is not conserved). As one can see, the difference is small. Moreover, experimentally one usually applies a gate voltage to tune  the Fermi level (or keep it constant), therefore in the following we will consider the case of a constant chemical potential. 

From Fig.~\ref{fig:Fig6}(a) follows that the conductivity for $\xi =10^{-5}$ cm
grows with increasing magnetic field.
Note, that the rate of the increase in conductivity with magnetic field (and accordingly the rate of the corresponding resistivity decrease) is larger than that for the relaxation time, see  Fig.~\ref{fig:Fig3} [except $B<10$ T in Fig.~\ref{fig:Fig3}(a)].  This tendency is a consequence of the suppression of spin-flip scattering with increasing $B$ and saturation of the number of states available for spin-conserving scattering (intra-Fermi-circle scattering) with increasing magnetic field. Thus, the scattering processes become more and more forward-scattering ones, so the  corresponding transport relaxation time becomes longer. Accordingly, the corresponding magnetoresistance is very small (negative or positive) in the regime of weak magnetic fields,  while for strong fields it is negative and relatively large,  Fig.~\ref{fig:Fig6}(b). 
In turn, for $\xi = 10^{-6}$ cm, the conductivity is much larger and decreases with increasing field until say 10 T (depending on the chemical potential $\mu$) and then increases again. The corresponding 
magnetoresistance is then very small and positive in the major part of the magnetic field region shown in Fig.~\ref{fig:Fig6} (c,d), and changes sign to negative only for strong  magnetic field, see Fig.~\ref{fig:Fig6}(d). 

To check the optimal values of the correlation length $\xi$, we show explicitly in  Fig.~\ref{fig:Fig7} the longitudinal conductivity and magnetoresistance as a function of the correlation length $\xi$ for two values of the external magnetic field. As follows from  this figure, the conductivity is large for $\xi =10^{-6}$ cm and drops rather quickly  when $\xi$ varies from $\xi =10^{-6}$ cm to  approximately $\xi \approx 5\times 10^{-6}$ cm, while the corresponding magnetoresistance remains then very small and  positive. 
 This is because scattering by Rashba fluctuations is relatively small for  $\xi \ll 5\times 10^{-6}$ cm.
In turn, when $\xi$ varies from $\xi \approx 5\times 10^{-6}$ cm  to  $\xi =10^{-5}$ cm, the conductivity is rather small, while magnetoresistance is negative and its absolute value increases with increasing $\xi$.   Thus, the optimal values of $\xi$ for having large negative magnetoresistance is between  $\xi =5\times 10^{-6}$ cm  and $\xi =10^{-5}$ cm.

\section{Summary and conclusions}

We have analysed transport properties of graphene in an in-plane external magnetic field, with randomly fluctuating Rashba spin-orbit interaction. Our main objective was to study the longitudinal magnetoresistance, and  the  
considerations were limited to low energy states in the vicinity of the Dirac points. To find the conductivity and then the longitudinal magnetoresistance, we used Kubo approach with the Green function formalism. We have calculated electron relaxation time due to scattering by Rashba field fluctuations and also included the vertex corrections. 

Numerical results based on the derived analytical formulas clearly show, that for the 
correlation length $\xi$ between  $\xi =5\times 10^{-6}$ cm  and $\xi =10^{-5}$ cm, the 
electrical conductivity of graphene with fluctuating Rashba interaction in external magnetic field applied in the graphene plane increases with increasing magnetic field. The corresponding resistivity becomes reduced by the external magnetic field, i.e. magnetoresistance is negative. This is in agreement with experimental observations~\cite{cite-key,PhysRevB.101.075425}.

\section*{acknowledgements}
This work has been supported by the National Science Center in Poland (NCN) as a research project No. UMO-2018/31/D/ST3/02351. The authors thank Evgeny Sherman for reading the manuscript and very useful comments. VD also thanks Volodymyr Ivanov for discussion 
%his contribution 
at the initial stage of this work.

\appendix

\begin{widetext}

\section{Calculation of the vertex function}
\label{App:A}

Here, we consider the equation for the vertex function defined in Eq.~\eqref{eq:Jvertex} and shown  graphically in Fig.~\ref{fig:Fig5}. This equation takes the form
\begin{eqnarray}
\label{eq:a1}
\Gamma _{k\sigma x} 
=\tau _x+\langle \lambda^{2}\rangle\int \frac{d^2{\bf k'}}{(2\pi )^2}\, F(|{\bf k-k'}|) %\hskip1cm
%\nonumber \\ 
%\times
\frac{\tau _y\, (\mu -\Delta _{\sigma }+v\btau \cdot {\bf k'})\; \Gamma _{k'\sigma x}} 
{(\mu -\varepsilon _{1\sigma }(k')-i\delta )\, (\mu -\varepsilon _{2\sigma }(k')-i\delta )}
%\nonumber \\ 
%\times
\frac{(\mu -\Delta _{\sigma }+v\btau \cdot {\bf k'})\, \tau _y}
{(\mu -\varepsilon _{1\sigma }(k')+i\delta )\, (\mu -\varepsilon _{2\sigma }(k')+i\delta )} 
\nonumber \\
+\langle \lambda ^{2}\rangle\int \frac{d^2{\bf k'}}{(2\pi )^2}\, F(|{\bf k-k'}|) % \hskip2cm 
%\nonumber \\ 
%\times
\frac{\tau _x\, (\mu -\Delta _{\bar{\sigma }}+v\btau \cdot {\bf k'})\; \Gamma _{k'\bar{\sigma }x}}
{(\mu -\varepsilon _{1\bar{\sigma }}(k')-i\delta )\, (\mu -\varepsilon _{2\bar{\sigma }}(k')-i\delta )}
%\nonumber \\ 
%\times
\frac{(\mu -\Delta _{\bar{\sigma }}+v\btau \cdot {\bf k'})\, \tau _x}
{(\mu -\varepsilon _{1\bar{\sigma }}(k')+i\delta )\, (\mu -\varepsilon _{2\bar{\sigma }}(k')+i\delta )}.\hspace{0.2cm} 
\end{eqnarray}
Let us assume that the vertex function $\Gamma _{k\sigma x}$ has the following matrix form
\begin{eqnarray}
\label{eq:a2}
\Gamma _{k\sigma x}=\tau _x\, g_{1\sigma k}+\tau _y\, g_{2\sigma k}+g_{3\sigma k}\, ,	
\end{eqnarray} 
where $g_{i\sigma k}$ are certain functions of $k$. 
Substituting \eqref{eq:a2} to \eqref{eq:a1} we obtain the following set of equations for the functions $g_{i\sigma k}$
%\ESc{Here and in the main text it would be better where possible to put common factor $v^{2}$ out of the parentheses. Also, integration elements in the momentum space are presented differently in these equations and others (here and in the main text).}
%\begin{widetext}
\begin{eqnarray}
\label{33}
g_{1\sigma k}=1+\langle \lambda^{2} \rangle \int \frac{d^2{\bf k'}}{(2\pi )^2}\, F(|{\bf k-k'}|)
\frac{-\mu _\sigma ^2g_{1\sigma k'}-v^2k_x'^2g_{1\sigma k'}-2v^2k_x'k_y'g_{2\sigma k'}+v^2k_y'^2g_{1\sigma k'}
-2\mu _\sigma vk_x'g_{3\sigma k'}} 
{(\mu _\sigma -vk'-i\delta )\, (\mu _\sigma +vk'-i\delta )\,
(\mu _\sigma -vk'+i\delta )\, (\mu _\sigma +vk'+i\delta )} 
\nonumber \\
-\langle \lambda^{2} \rangle \int \frac{d^2{\bf k'}}{(2\pi )^2}\, F(|{\bf k-k'}|)
\frac{-\mu _{\bar{\sigma }}^2 g_{1\bar{\sigma }k'}-v^2k_x'^2 g_{1\bar{\sigma }k'}
-2v^2k'_xk'_yg_{2\bar{\sigma }k'}+v^2k_y'^2g_{1\bar{\sigma }k'}-2\mu _{\bar{\sigma }}vk'_xg_{3\bar{\sigma }k'}}
{(\mu _\sigma -vk'-i\delta )\, (\mu _\sigma +vk'-i\delta )\,
(\mu _\sigma -vk'+i\delta )\, (\mu _\sigma +vk'+i\delta )} \; ,
\end{eqnarray}	
\begin{eqnarray}
\label{34}
g_{2\sigma k}=\langle \lambda^{2} \rangle \int \frac{d^2{\bf k'}}{(2\pi )^2}\, F(|{\bf k-k'}|)\, 
\frac{\mu _\sigma ^2g_{2\sigma k'}-v^2k_x'^2g_{2\sigma k'}+2v^2k_x'k_y'g_{1\sigma k'}+v^2k_y'^2g_{2\sigma k'}
+2\mu _\sigma vk_y'g_{3\sigma k'}} 
{(\mu _\sigma -vk'-i\delta )\, (\mu _\sigma +vk'-i\delta )\,
(\mu _\sigma -vk'+i\delta )\, (\mu _\sigma +vk'+i\delta )} 
\nonumber \\
-\langle \lambda^{2} \rangle \int \frac{d^2{\bf k'}}{(2\pi )^2}\, F(|{\bf k-k'}|)\, 
\frac{\mu _{\bar{\sigma }}^2 g_{2\bar{\sigma }k'}-v^2k_x'^2 g_{2\bar{\sigma }k'}
+2v^2k'_xk'_yg_{1\bar{\sigma }k'}+v^2k_y'^2g_{2\bar{\sigma }k'}+2\mu _{\bar{\sigma }}vk'_yg_{3\bar{\sigma }k'}}
{(\mu _\sigma -vk'-i\delta )\, (\mu _\sigma +vk'-i\delta )\,
(\mu _\sigma -vk'+i\delta )\, (\mu _\sigma +vk'+i\delta )} \; ,
\end{eqnarray}	
\begin{eqnarray}
\label{35}
g_{3\sigma k}=\langle \lambda^{2} \rangle \int \frac{d^2{\bf k'}}{(2\pi )^2}\, F(|{\bf k-k'}|)\, 
\frac{2\mu _\sigma v(k_x'g_{1\sigma k'}+k_y'g_{2\sigma k'})+(\mu _\sigma ^2+v^2k'^2)g_{3\sigma k'}} 
{(\mu _\sigma -vk'-i\delta )\, (\mu _\sigma +vk'-i\delta )\,
(\mu _\sigma -vk'+i\delta )\, (\mu _\sigma +vk'+i\delta )} 
\nonumber \\
-\langle \lambda^{2} \rangle \int \frac{d^2{\bf k'}}{(2\pi )^2}\, F(|{\bf k-k'}|)\, 
\frac{2\mu _{\bar{\sigma }}v(k_x'g_{1\bar{\sigma }k'}+k_y'g_{2\bar{\sigma }k'})
+(\mu _{\bar{\sigma }} ^2+v^2k'^2)g_{3\bar{\sigma } k'}}
{(\mu _\sigma -vk'-i\delta )\, (\mu _\sigma +vk'-i\delta )\,
(\mu _\sigma -vk'+i\delta )\, (\mu _\sigma +vk'+i\delta )} \; .
\end{eqnarray}	
%\end{widetext}
%
Six linear algebraic equations (54)-(56) for $g_{1\uparrow ,\downarrow },g_{2\uparrow ,\downarrow },g_{3\uparrow ,\downarrow }$ should be solved numerically as a function of angle $\phi $.
In the integrals, one can write 
\begin{eqnarray}
\label{37}
&&k_i'=n_i ({\bf k}\cdot {\bf k'})/k=n_ik'\cos \theta ,
\nonumber \\
&&k_ik_j=n_in_j\, k'^2,	
\end{eqnarray} 
where $n_i=k_i/k$ and $\theta $ is the angle between vectors ${\bf k}$ and ${\bf k'}$. In addition the integration over ${\bf k'}$ with the function $F(|{\bf k-k'}|)$ can be presented as
\begin{eqnarray}
\label{38}
\int \frac{d^2{\bf k}'}{(2\pi )^2}\, F(|{\bf k-k'}|)\, ...\; 
\delta (k'-\mu _\sigma /v) 
%\nonumber \\ 
=\frac{k_\sigma \xi ^2}{2\pi ^2}\int _0^\pi d\theta \; e^{-\xi ^2(k^2+k_\sigma ^2-2kk_\sigma \cos \theta )}...\, .
\end{eqnarray}
Consequently using Eqs.~(A3)-(A5), and taking $k=k_\sigma $, $\mu _\sigma =vk_\sigma $ we get a set of linear algebraic equations for $g_{i\sigma k}$
%\begin{widetext}

\begin{eqnarray}
\label{39}
g_{1\sigma k_\sigma }
=1+\frac{k_\sigma \xi ^2\langle \lambda^2\rangle }{4v\gamma _{1\sigma }}
e^{-2\xi ^2k_\sigma ^2}
\big[ n_y^2\, I_0(2\xi ^2k^2_\sigma )\, g_{1\sigma k_\sigma }
-n_xn_y\, I_0(2\xi ^2k^2_\sigma )\, g_{2\sigma k_\sigma }
- n_x\, I_1(2\xi ^2k^2_\sigma )\, g_{3\sigma k_\sigma }\big] \hskip1cm
\nonumber \\
+\frac{k_{\bar{\sigma }}\xi ^2\langle \lambda^2\rangle }{4v\gamma _{1{\bar{\sigma }}}}
e^{-\xi ^2(k_\sigma ^2+k_{\bar{\sigma }}^2)}
\big[ n_x^2\, I_0(2\xi ^2k_\sigma k_{\bar{\sigma }})\, g_{1\bar{\sigma }k_{\bar{\sigma }}}
%\nonumber \\
+n_xn_y\, I_0(2\xi ^2kk_{\bar{\sigma }})\, g_{2\bar{\sigma }k_{\bar{\sigma }}}
+n_x\, I_1(2\xi ^2k_\sigma k_{\bar{\sigma }})\, 
g_{3\bar{\sigma }k_{\bar{\sigma }}}\big] ,
\end{eqnarray}	

\begin{eqnarray}
\label{40}
g_{2\sigma k_\sigma }=\frac{k_\sigma \xi ^2\langle \lambda^2\rangle }{4v\gamma _{1\sigma }}
e^{-2\xi ^2k_\sigma ^2}
\big[ n_y^2\, I_0(2\xi ^2k^2_\sigma )\, g_{2\sigma k_\sigma }
+n_xn_y\, I_0(2\xi ^2k^2_\sigma )\, g_{1\sigma k_\sigma }
+n_y\, I_1(2\xi ^2k^2_\sigma )\, g_{3\sigma k_\sigma }\big]  \hskip1cm
\nonumber \\
-\frac{k_{\bar{\sigma }}\xi ^2\langle \lambda^2\rangle }{4v\gamma _{1{\bar{\sigma }}}}
e^{-\xi ^2(k_\sigma ^2+k_{\bar{\sigma }}^2)}
\big[ n_y^2\, I_0(2\xi ^2k_\sigma k_{\bar{\sigma }})\, g_{2\bar{\sigma }k_{\bar{\sigma }}}
%\nonumber \\
+n_xn_y\, I_0(2\xi ^2k_\sigma k_{\bar{\sigma }})\, g_{1\bar{\sigma }k_{\bar{\sigma }}}
+n_y\, I_1(2\xi ^2k_\sigma k_{\bar{\sigma }})\, g_{3\bar{\sigma }k_{\bar{\sigma }}}\big] .
\end{eqnarray}	

%\ESc{It is better for readability to put common factors $I_{0}$ out of the parentheses.}

\begin{eqnarray}
	\label{41}
	g_{3\sigma k_\sigma }=\frac{k_\sigma \xi ^2\langle \lambda^2\rangle }{4v\gamma _{1\sigma }}
	e^{-2\xi^{2}k_{\sigma}^2}
	\big[ \, I_1(2\xi^{2}k_{\sigma}^{2})\, (n_x\, g_{1\sigma k_\sigma }+n_y\, g_{2\sigma k_\sigma })
	+I_0(2\xi ^2k^2_\sigma )\, g_{3\sigma k_\sigma }\big]  
	\hskip1cm
	\nonumber \\
	-\frac{k_{\bar{\sigma }}\xi ^2\langle \lambda^2\rangle }{4v\gamma _{1{\bar{\sigma }}}}
	e^{-\xi ^2(k_\sigma ^2+k_{\bar{\sigma }}^2)}
	\big[ \, I_1(2\xi ^2k_\sigma k_{\bar{\sigma }})\, (n_x\, g_{1\bar{\sigma }k_{\bar{\sigma }}}
	+n_y\, g_{2\bar{\sigma }k_{\bar{\sigma }}}) 
	+I_0(2\xi ^2k_\sigma k_{\bar{\sigma }})\, g_{3\bar{\sigma } k_{\bar{\sigma }}}\big] ,
\end{eqnarray}	
where $n_x=k_x/k$ and $n_y=k_y/k$.
\end{widetext}

%=============================
% BIBLIOGRAPHY
%=============================
%\nocite{*}
\bibliography{bib.bib}

\end{document}